\documentclass{article}
\usepackage{cite} 
\usepackage[utf8]{inputenc}
\usepackage[left=3.1cm,right=6.1cm,top=2.1cm,bottom=3.1cm,marginparwidth=5.65cm]{geometry}
\usepackage{graphicx}
\usepackage{amsmath}
\usepackage{csquotes}
\usepackage{chemformula}
\usepackage{bbding}
\usepackage{marginnote}
\usepackage{comment}
\usepackage{bibunits}
\usepackage{enumitem}
\usepackage[hang,flushmargin]{footmisc}
\usepackage{tocloft}
\usepackage{titlesec}
\usepackage{titletoc}
\usepackage[hypertexnames=false]{hyperref}
\hypersetup{colorlinks=true, %
citecolor=blue,%
linkcolor= blue
}

\titlespacing{\section} {0pt}{1.5ex plus .1ex minus .2ex}{0pt}
\titlespacing{\subsection} {0pt}{1.5ex plus .1ex minus .2ex}{0pt}
\titlespacing{\subsubsection} {0pt}{1.5ex plus .1ex minus .2ex}{0pt}

\titlecontents{subsection}[2cm]{}{\contentslabel{2.2em}}{\hspace*{3.2em}}{\hspace*{0.15cm}\cftdotfill{1}{.}\contentspage}[]
\titlecontents{subsubsection}[3.2cm]{}{\contentslabel{3em}}{\hspace*{3.2em}}{\hspace*{0.15cm}\cftdotfill{1}{.}\contentspage}[]
\titlecontents{paragraph}[4.7cm]{}{\contentslabel{3.7em}}{\hspace*{3.2em}}{\hspace*{0.15cm}\cftdotfill{1}{.}\contentspage}[]

\begin{document}

\vspace*{2cm}

\begin{center}
\Large
Translation of a 1971 paper by Otto Rössler\\
with added margin notes and a commentary\\
\end{center}

\vspace*{0.2cm}

\begin{center}
{\large Walter Fontana and Philipp Honegger}\\
\vspace*{0.2cm}
{\normalsize Systems Biology, Harvard Medical School, Boston MA 02115 USA}
\end{center}

\vspace*{1cm}

\noindent \emph{Some time ago we came across a rather unknown and, in our opinion, intriguing paper by Otto R\"ossler on autocatalysis. That the paper is written in German might be one reason for having spent 50 years below the radar despite being of potential interest to many people in the field (as we found anecdotally). We felt it is time to translate it. We provide some margin notes and a commentary/introduction that has been solicited for a book project aimed at a general scientific audience.}

\vspace*{1cm}

{
\tableofcontents
}

\newpage
\phantomsection 
\addcontentsline{toc}{section}{A System Theoretic Model of Biogenesis (by Otto Rössler)}

\begin{center}
{\Large A System Theoretic Model of Biogenesis$^{\ast}$}
\let\thefootnote\relax\footnotetext{\noindent $^{\ast}$Translated from German by 
Philipp Honegger and Walter Fontana\\
Systems Biology, Harvard Medical School, Boston MA 02115}\\[0.5cm]
{\large \bf Otto E. R\"ossler}\\[0.2cm]
{\large Theoretical Chemistry, University of T\"ubingen}\\[0.2cm]
originally published in\\ 
\emph{Zeitschrift f\"ur Naturforschung} \textbf{26 b}, 741--746 [1971]\\
(received March 15th 1971)\\
\end{center}

\vspace*{3cm}

\begin{center}
{\bf \large Abstract}
\end{center}

\noindent Three types of abstract chemical reaction systems are described: 1.~The generalized catalytic system, 2.~the generalized autocatalytic system, 3.~a spontaneously evolving chemical system. The significance of the second and third system for a very early phase of pre-biological evolution is discussed.

\newpage
\begin{bibunit}[ieeetr]

We will be concerned not with a system-theoretic process that can explain the origin of the first living system, but only with a process that can explain the origin of an autonomously growing chemical system sharing a certain property with a living system. That property consists in the capacity to become, through a mechanism of stepwise modification, the \enquote{ancestor} of an almost unlimited multitude of diverse, autonomously growing chemical systems.

Such a process is of interest only if it is simple, i.e.\@ if it is based on a single easily describable principle, and if it does not require extraordinary chemical preconditions.

This demand is not necessarily unrealistic. In technology it is customary to synthesize systems with rather extraordinary properties using simple building blocks while following simple principles. By analogy, the combination of simple chemical reactions can result in chemical systems with properties that are new and unusual compared to those of their components. It is even possible for chemical systems of this kind to form spontaneously.

This fundamental possibility shall first be demonstrated using an illustrative model. The properties of that system, in particular those of an important special case, will subsequently be needed in specifying the system proper.

\begin{center}
{\bf\emph{An example}}
\end{center}

It is known that under certain conditions some chemical substances can, upon addition of another substance, spontaneously react with it to generate one or more new substances. This is represented schematically as \ch{A + B -> C + D}, or equivalently:

\begin{center}
    \includegraphics[width=0.22\textwidth]{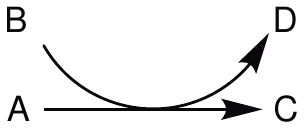}
\end{center}

\noindent (We assume irreversibility for the sake of simplicity.) Sometimes, the product will spontaneously give rise to further products:

\begin{center}
    \includegraphics[width=0.62\textwidth]{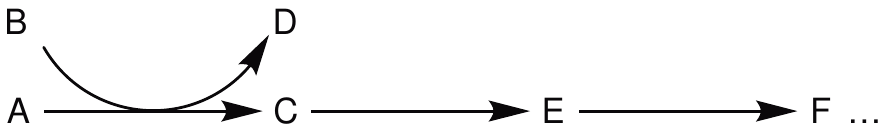}
\end{center}

\noindent However, one might also add a new substance \ch{E} to the product \ch{C} of the first reaction, causing the spontaneous formation of further products, and so on:

\begin{center}
    \includegraphics[width=0.45\textwidth]{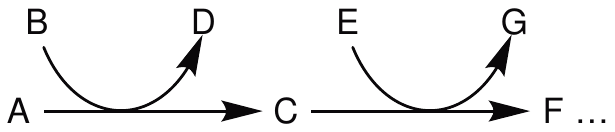}
\end{center}

\begin{figure}[h]
    \centering
    \includegraphics[width=0.475\textwidth]{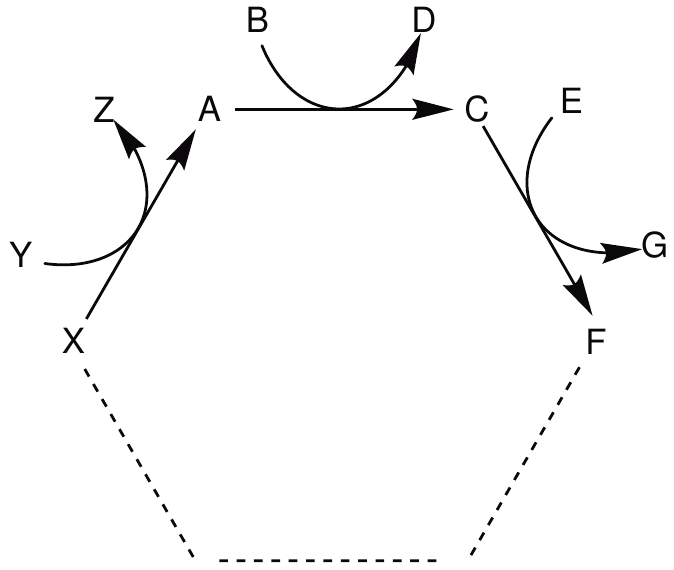}
    \caption{Kinetic scheme of a generalized catalytic reaction (principle).}
    \label{fig:abb_1}
\end{figure}

Thus, there are many degrees of freedom in the design of reaction sequences. In some sequences a subset of atoms from the initial substance \ch{A} is conserved over many steps, while in others none of the downstream products (that are being considered) has a single atom in common with upstream reactants. Because of these degrees of freedom, it is fundamentally possible to find sequences that form a cycle after a few (say $n$) steps (Figure \ref{fig:abb_1}).

In the simplest case, following the principles of stoichiometry, a new molecule of \ch{A} is generated after $n$ steps for each molecule of \ch{A} consumed in the cycle to form \ch{C}. Thus, in final analysis, the substance \ch{A} is not depleted by its reaction with \ch{B} to form \ch{C}---a reaction enabled by the very presence of \ch{A}---but rather behaves like a catalyst that is continuously regenerated. The same is true for substances \ch{C}, \ch{F}, ..., \ch{X}. The reaction system in Figure \ref{fig:abb_1} can therefore be called a \enquote{\textit{generalized catalytic reaction}}. Substances \ch{D}, \ch{G}, ..., \ch{Z} constitute the \enquote{products of catalysis} in this reaction. The \enquote{substrate} from which these products are generated consists of the externally added substances \ch{B}, \ch{E}, ..., \ch{Y}. The set of these substances cannot be empty, or the cycle would only consist of pure follow-up reactions, but must include at least one food substance. Each of the \enquote{generalized catalysts} \ch{C}, \ch{F}, ..., \ch{X} is equivalent. A single one of them, when added to the set of food substances, is sufficient to generate all products. Indeed, from its mass all other generalized catalysts that participate in the cycle are formed.

This is an example for how very simple chemical prerequisites can result in an unexpected (systemic) effect: \textit{Catalysis without the presence of a catalyst} (understood in a narrow sense). \marginnote[annotation1]{(i)} It seems possible to assemble reaction sequences of this kind artificially. If all that matters is the creation of a particular product, many degrees of freedom exist to achieve the eventual regeneration of the substance initiating the reaction sequence. Most importantly, the generalized catalytic reaction is an example for how a reactive system with non-trivial behavior can emerge \enquote{by itself} under suitable conditions.

\begin{center}
{\bf\emph{A special case of generalized catalysis}}
\end{center}

\begin{figure}[h]
    \centering
    \includegraphics[width=0.475\textwidth]{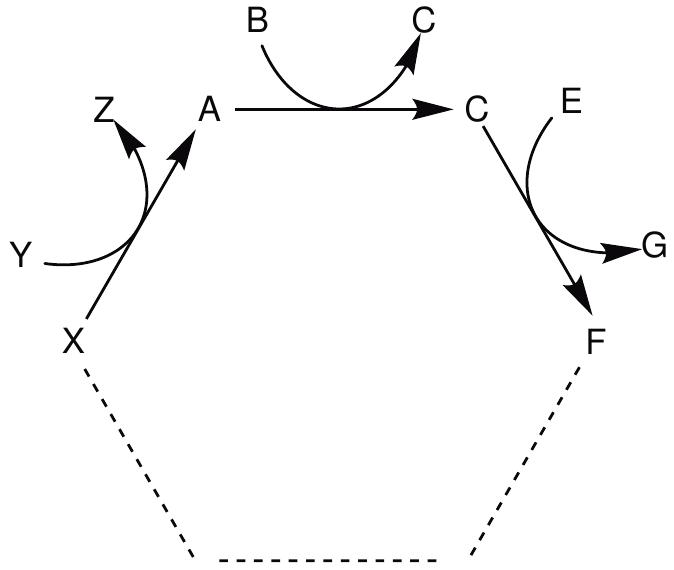}
    \caption{Kinetic scheme of a generalized autocatalytic reaction (principle).}
    \label{fig:abb_2}
\end{figure}

A particular case of the scheme in Figure \ref{fig:abb_1} deserves special attention: Once the stoichiometric coefficient of the net reaction \ch[arrow-offset=0.5em]{A->A} (and \ch[arrow-offset=0.5em]{C->C}, etc.) becomes greater than 1, that is, a more than complete \enquote{regeneration} of the substances participating in the cycle occurs, the system-theoretic phenomenon of \textit{autonomous growth} can ensue. As shown in Figure \ref{fig:abb_2}, this excess regeneration occurs already if the stoichiometric coefficient of 2 occurs in just one of the reaction steps (\ch[arrow-offset=0.5em]{A+B->C+C+D}), \marginnote[annotation2]{(ii)} that is, when forming two equal groups of atoms. The phenomenon of \enquote{generalized autocatalysis} occurs as long as the molecules produced in excess encounter the same reaction conditions as their predecessors. (Note that this proof does not rely on specific kinetic arguments, but on general relational ones. However, consideration of the details in each case permits additional, more specific assertions. The type of reaction and the rate constants in the forward and reverse direction determine the particular growth function, the rate of growth and the range of growth; allowing for inhomogeneous conditions enables spatial differentiation; etc.\@ \cite{ref_1}).

One difference between true \enquote{autocatalysis} and the more easily realizable so-called cross-catalysis (two catalysts that mutually produce each other) is that no catalyst in the narrow sense is needed. Nonetheless, the case of autocatalysis is covered as a special case in the scheme of Figure \ref{fig:abb_2} (by using 2 reaction steps and just one input substance) if one assumes a simple Michaelis-Menten-type reaction mechanism. Cross-catalysis and cyclic catalyses of higher order yielding autonomous growth are also included as special cases, if one considers that the cycle may contain additional cross-links.

Just as with generalized catalysis, the cycle can again be reconstituted in the presence of the food substances from a single substance of those that constitute the cycle. In this case, however, a single molecule suffices in principle \cite{ref_2} Thus, the generalized autocatalytic reaction system demonstrates the principles stipulated at the outset with particular clarity.

\begin{center}
{\bf\emph{Two concluding remarks on the topic of generalized autocatalysis}}
\end{center}

It must be mentioned that the system shown in Figure \ref{fig:abb_2} is by no means unknown in the literature. It has been discussed in the context of \enquote{isothermal chain reactions} \cite{ref_3} and was mentioned repeatedly in a purely theoretical context \cite{ref_4}. However, it has not yet been expressed as an abstract principle of relevance to the artificial and spontaneous synthesis of systems.

Finally, as always in a systems-theoretic context of chemistry, we must discuss the thermodynamic \enquote{admissibility} of a proposed \enquote{kinetics}. A proof for the admissibility of \enquote{autocatalysis} in the general sense of a substance promoting its own formation has already been provided within the framework of the thermodynamics of irreversible processes (extended to strongly nonlinear dependencies between fluxes and forces) \cite{ref_5}.

\begin{center}
{\bf\large Assumptions of the model}
\end{center}

The behavior of autonomous growth derived in the previous example belongs to the simplest patterns of systemic behavior producible by a proper \textit{interconnection} of simple reactions. One can expect to also find examples that do not require catalysts in the narrow sense for the other well-known behavior patterns in reaction systems (such as oscillations and multi-stability). \marginnote[annotation2]{(iii)} However, in the present context the issue is another one.

The plugging together of systems of a certain type into \enquote{systems of higher-order} is a common construction principle in technology. The subsystems maintain thereby a certain level of autonomy. The same principle is also found in biology. It is not far fetched to hypothesize that chemical systems of higher order are possible as well.

This is the case for the process of a systemic chemical evolution that will be described next. It can be characterized as \enquote{second-order autonomous growth}. The difference to ordinary autonomous growth consists in the autocatalytic generation not of the number of molecules but the number of systems. An aspect in common with ordinary autonomous growth consists in the spontaneous emergence of such behavior.

The preconditions for such a process are simple. They are included in the assumptions made for terrestrial \enquote{chemical evolution}\cite{ref_6}, but they are so general that they could be satisfied by completely different chemical conditions.

The most essential of these preconditions is identical to the principle at the basis of organic chemistry. It is the fact that with carbon it is possible to generate an almost unlimited number of distinct products using a small number of suitable starting materials.\cite{ref_7} It is conceivable, however, that under different physical conditions other kinds of starting materials (without carbon) also possess the capability of constructing a universal library of products, for instance by forming alternating chains.

The derivation of further preconditions is best approached through the concept of a network of possible reactions. This is the set of all reactions that become possible by positing a few types of substances. This network is very large if by \enquote{possible} we mean the energetically (thermodynamically) feasible reactions. For instance, just the two substances \ch{CO2} and \ch{H2O} yield together with an energy source (in the simplest case: light quanta of a particular energy) at atmospheric pressure and room temperature a universal library of energetically possible products and their associated reactions.

The network of energetically possible reactions is a very theoretical concept. It includes the more interesting network of \textit{spontaneously} possible reactions, that is, those reactions that are not only possible energetically (due to a gradient in free enthalpy) but also \enquote{kinetically} because the activation energies required to form transition states of the Eyring type are sufficiently low at the given temperature. This second network is usually much smaller (as in the example just mentioned). However, a connection between these two concepts will be shown next.

\marginnote[annotation4]{(iv)} The network of reactions that are energetically but not necessarily kinetically possible contains a very large number of cyclic reaction systems of the kind shown in Fig.~\ref{fig:abb_2}. The reasons are twofold: First, a product of the $n^\text{th}$ reaction step does not need to be more similar to the starting substance than any arbitrary substance.\cite{ref_8} Second, the probability that a substance picked randomly from the network can react with any other substance in the network is, on average, constant. (This changes only if more is known about one of the substances, for example that it is energy-rich, or if both substances are specified). The principle of a random wiring of the network of energetically possible reactions follows from the possibility of a molecule containing the most varied atom groups capable of the most varied reaction possibilities.

Among the very large number of cyclical reaction systems contained in the network of energetically but not necessarily kinetically possible reactions there is a still very large number of cycles all of whose steps are also kinetically possible. Although these cycles are connected to the starting substances by energetically feasible reactions (that is, reactions with a nonzero rate constant), none of them can self-start within an arbitrarily large amount of time [even] when the starting substances are connected to an inexhaustible reservoir. The reason is the small magnitude of the rate constants of the reaction steps leading to these cycles.

\marginnote[annotation3]{(v)} This statement contradicts the assumption of a continuum kinetics. Under that assumption, a cycle of the kind described would start up already with infinitesimally small concentrations. Moreover, due to the small rate constants, the whole reaction system would be coupled to an extent that makes it meaningless to speak of individual subsystems. Yet, the above statement follows from the fact that, say, a production rate of 10$^{-12}$ molecules per liter and minute in a reaction volume of 1 liter does \emph{not} mean that the hypothetical 10$^{-12}$ molecules produced after a minute are instantly available to react with other molecules. Concentrations below 1 molecule (relative to the total homogeneous reaction volume) have the same effect as the introduction of a \emph{time delay} in an otherwise continuous dynamical system.\cite{ref_9} This fact is one of the basic tenets of any process of chemical evolution.

To start up one of the possible spontaneous cycles in the network requires an uninterrupted chain of possible spontaneous reactions leading to each of the substrates on which the cycle depends. This \marginnote[annotation4]{(vii)} requirement seems more easily satisfiable for one of the generalized catalysts than for the entirety of input substrates. Furthermore, the latter require a continuous supply. The same set of input substrates, however, is often sufficient to support an entire universal library of autonomously growing systems. \marginnote[annotation4]{(vi)} Therefore, a very large number of more or less equivalent sets of input substrates exist that make the requirement for the spontaneous formation of a single one of them no longer that unrealistic.

\begin{center}
{\bf\large The system-theoretical model}
\end{center}

Having specified the essential preconditions, assume next that a small number of starting materials, which are kept at constant concentration, fulfill the requirements for spontaneously setting off a single simple cycle. By virtue of its catalytic function, this cycle then produces a whole series of substances, to wit, in addition to the other catalysts in the cycle, also all products of the cycle (C, G, ..., Z in Fig.~\ref{fig:abb_2}), which can further react with one another and with the starting materials. \marginnote[annotation5]{(viii)} At the end of the growth phase of the cycle, the primary products are available at steady-state concentrations, that is, they will also be replenished at concentrations comparable to those of the given starting substances.

Thus, the special case that all requirements for the realization of one cycle are satisfied has the same effect as if a larger set of starting substances had been provided (some of them at constant concentration). Since the probability of jump starting one of the many possible spontaneous cycles in the network of possible reactions increases stronger than linear (in fact combinatorially) in the number of substances present, the starting of one cycle facilitates the occurrence of a second one, which in turn facilitates the occurrence of a third, and so on, to an ever increasing extent. This evidently points at the existence of a \emph{threshold}: After starting up a critical number of cycles, the probability of kicking off at least one further cycle, and from there on ever more cycles, suddenly becomes almost 1.

The same statement can be represented in simple formal terms thus: 
\marginnote[annotation5]{(ix)} 
\begin{eqnarray} 
 n &= &f_1(N) \\
 N &= &f_2(n), \nonumber
\end{eqnarray}
where $n$ is the number of substances present in the reaction network, $N$ is the number of already active cycles, and $f_1$ and $f_2$ are two functions increasing monotonically with $N$ and $n$, respectively. This positive feedback becomes \enquote{super-critical} only if the product of $f_1$ and $f_2$ exceeds 1.

The value of the threshold depends on special conditions other than on $n_0$ (the number of externally added substances kept partially at constant concentration). If, for example, the initial mixture contains by chance, in addition to the substances needed for a universal library, other materials that possess a small yet broad catalytic activity, such as metal ions, then the threshold would be lowered considerably.

With this, the process sought has been found. One could call this behavior, which, after crossing a threshold, spontaneously generates ever more autonomously growing systems an \enquote{autonomous growth of second order}. It produces not only ever more autonomously growing systems, but also systems whose immediate spontaneous formation is increasingly unlikely. \marginnote[annotation6]{(x)} While simple autocatalytic systems, such as depicted in Fig.~\ref{fig:abb_2}, are the only ones growing at first, they are soon joined by interconnected ones and others that contain \enquote{true} catalysts. Finally, autonomously growing systems of very high complexity can become activated.

Where the threshold resides in each specific case, and whether it can be overcome within a natural system of reactions under natural conditions and on a natural time scale, is a question that needs to be studied separately. Likewise, one has to reckon with the emergence of other systemic behaviors of the second order. The behavior described here is only valid as long as the coupling of individual unstable systems is weak. It seems as if the inclusion of diffusion could delay the phase of a general interconnection to an arbitrary extent. The description of the processes following the \enquote{explosive phase} is, however, of secondary interest, much as in the case of the first example.\cite{ref_10a}

So far the emphasis was on the explosion-like aspect of the behavior. Another, equally important, aspect is the automatic \enquote{screening} of a universal library. This behavior is also known from biology, where it relates to the far more special universal library of possible nucleic acid sequences. Summing up, the hypothesis proposed here consists in positing that the general process just described is suited to start a specific, for example a biological, one. \marginnote[annotation6]{(xi)} This would be the case if among the screened systems there was one that could become the starting point of an explosion of the second-order functioning no longer blindly, but in accordance with a principle of descent (and thus under certain additional conditions far more effective).

\begin{center}
{\bf\large Discussion}
\end{center}

The two essential results that were achieved are 1.~the specification of the structure of a generalized autocatalytic system and 2.~the structure of a \enquote{second-order autocatalytic system} representing an example of chemical evolution.

Both results were not achieved with kinetic or dynamic methods. Rather, the derivation followed from simple relational schemes (in the first case: from A follows B, stoichiometry, relations of consumption. In the second case: cycle, network, wiring). The restriction to this general level loses a great deal of specific information. Yet, it allows certain law-like regularities to become visible. For example, the proof that all chemical systems of the type shown in Fig.~\ref{fig:abb_2} are unstable cannot be achieved by kinetic methods, that is, using the theory of dynamical systems. Even \marginnote[annotation6]{(xii)} when limiting the analysis to the special case of linear reaction systems described by first-order ordinary differential equations, the proof that such systems exhibit autonomous growth, that is, that they possess a critical point at the origin of phase space or, equivalently, a positive eigenvalue of their characteristic polynomial, presents considerable difficulty. This is even more the case for chemical systems of this type that must be described by nonlinear and partial differential equations. Likewise, the principle of second-order instability cannot be derived in general at the level of an explicit dynamic description.

The results achieved thus suffer from the disadvantage that they do not completely describe any specific case. However, they enable specific quantitative models that could not be constructed otherwise. For instance, it is possible to model the unstable system of the second kind at the level of cycles as a system of switches with conditional connections and weak initial conditions (that is, as a nonlinear finite automaton). A simulation at the level of substances is also possible, either also in terms of switches (as a rough qualitative model) or more finely grained by using system of functional differential equations with a weakly inhomogeneous initial condition. The most complicated case would be represented by a dynamical model consisting of partial functional differential equations. With the help of quantitative models an assessment can be made about the extent to which natural reaction systems fulfill the conditions of the model.

In conclusion it is important to note that the described model, which represents a very general form of a chemical evolutionary theory, can be of significance for \emph{terrestrial} biogenesis, if certain minimal requirements regarding the functional complexity of the biological \enquote{ancestral system} must be made.\cite{ref_11}

\vspace*{0.5cm}

\noindent
I'm grateful to the Deutsche Forschungsgemeinschaft for supporting this work.

\vspace*{0.4cm}

\phantomsection
\addcontentsline{toc}{section}{References}
\renewcommand{\refname}{}
\subsection*{References}
\putbib[refs2]
\end{bibunit}

\subsection*{Margin notes by P.H.~and W.F.~to the translated paper}
\vspace*{0.4cm}

\renewcommand{\refname}{}
\begin{bibunit}[ieeetr]

\begin{enumerate}[label=(\roman*)]
\item
By \enquote{catalysis in a narrow sense} R\"ossler means a \textit{net} reaction that proceeds by a catalytic mechanism: \ch[arrow-offset=0.5em]{A + C -> B + C}. In contrast, \enquote{catalysis without catalyst} refers to a combination of non-catalytic net reactions whose overall effect is catalytic, such as \ch[arrow-offset=0.5em]{A + B -> C}, \ch[arrow-offset=0.5em]{C -> D}, \ch[arrow-offset=0.5em]{D -> A + E}, which consumes B and produces E. This is known today as \textit{network catalysis}.
\item
The additional instance of C does not have to be produced in the same step.
\item
Indeed, autocatalytic networks (sensu R\"ossler) can exhibit all of these behaviors, as was shown experimentally.\cite{whi16a}
\item
Lacking any chemical detail in the specification, the reaction network is treated as more or less uniform. In reality, chemical networks are anisotropic as they consist of highly reactive molecules, moderately reactive and relatively inert ones. Groups of atoms that often occur together in molecules (i.e.\@ functional groups) entertain specific relations of reactivity that can lead to dense subnetworks (e.g.\@ formose chemistry) and sparse ones (e.g.\@ noble gases).
\item
R\"ossler makes the case for stochastic chemical kinetics. When the number of molecules is large, the height of a barrier determines the rate at which molecules flow over it. When the number of molecules is small, flow rates must be replaced with transition probabilities; low probabilities (high barriers) imply longer average waiting times to a crossing event.
\item
While an uninterrupted reaction sequence is required for each substrate feeding into a cycle, only one is required for the cycle itself.
\item
This harks back to the idealization of a uniform reaction network three paragraphs ago.
\item
While R\"osslers argument of an increasing occupation of chemical space has merit, it neglects that reaction branches can peter out in inactive materials. Autocatalysis is \enquote{self-healing} as long as its growth rate outpaces its decay rate.\cite{bla09b}
\item
Network autocatalysis increases the number of replenished molecular species and, typically, the number of reactions increases more than linearly with the number of molecular species in a network.
\item
\enquote{True} here refers to single-molecule as opposed to network catalysis.
\item
R\"ossler's general argument is based on statistical properties of a roughly uniform reaction network. Without further details it is difficult to assess its suitability as a model of biogenesis. But the idea that in some corner of chemistry the process laid out here can occur is intriguing.
\item
This statement seems puzzling, if interpreted from the vantage point of deterministic dynamics. However, at very low concentrations, the discrete nature of particles requires a stochastic description. The system can die out by fluctuations even if particle numbers are initially positive and expectation values are increasing. The zero point is therefore at best a conditionally unstable point.
\end{enumerate}

\subsection*{Margin note references}
\putbib[refs2]
\end{bibunit}
\newpage
\phantomsection 
\addcontentsline{toc}{section}{The Wheels of Chemistry}

\begin{center}
\Large
The Wheels of Chemistry\\
\end{center}

\vspace*{0.2cm}

\begin{center}
{\large Walter Fontana and Philipp Honegger}\\
\vspace*{0.2cm}
{\normalsize Systems Biology, Harvard Medical School, Boston MA 02115 USA}
\end{center}

\vspace*{0.3cm}

\begin{bibunit}[ieeetr]

Otto R\"ossler's 1971 paper \cite{Rossler1971} is a landmark in the landscape of systems chemistry that was all but missed despite providing a consequential refinement of the idea of generalized autocatalysis, which refers to a system that facilitates its own growth. R\"ossler is better known for the chaotic attractor named after him \cite{Rossler1976} than for the principle of \enquote{chemical space} exploration that he outlines in his paper.

One problem with R\"ossler's paper is obvious: it is written in German. We therefore provide a translation (with a few margin notes) to at least remove that obstacle. Besides language, another factor that might have reduced its visibility was the publication, earlier in the same year, of Manfred Eigen's work on \enquote{Selforganization of Matter and the Evolution of Biological Macromolecules} \cite{Eigen1971}, which also deals with autocatalysis. Eigen's paper quickly became a classic. Again in 1971, Stuart Kauffman started his thread on \enquote{autocatalytic sets} in a long appendix piggybacking on his introduction of random Boolean networks \cite{Kauffman1971}. The chemical engineer Tibor G\'anti was among those who characterized living systems through an abstract chemical model of several interacting autocatalytic subsystems. He published (in Hungarian) his chemical automaton---the \enquote{chemoton}---in, you guessed it, 1971 \cite{ganti71}. (For an English-language reference, see \cite{Ganti1975}.) G\'anti, however, is less concerned with a process of chemical evolution towards a living state than with the specification of a working model in support of a conceptual and computational exploration of that state. 1971 appears to have been the \emph{annus mirabilis} of autocatalysis. 

While vastly differing in the level of detail, the three papers by Eigen, Kauffman, and R\"ossler constitute mutually opposite corners of a triangle of viewpoints informing the origin of life. R\"ossler begins with the discussion of emergent phenomena in chemistry, specifically catalysis. The traditional view of a catalyzed reaction considers a molecule that forms short-lived intermediates with reactants to open up a more favorable path on the free energy surface of their chemical transformation into products. In the process, the catalytic molecule is returned to its original chemical state and therefore shows up as the same entity on both sides of the overall reaction, \ch{A + B + C -> D + E + C}. The catalyst \ch{C} cannot be canceled, since the reaction mechanism depends on its presence. Catalysis, however, can also be achieved by constructing a cycle of consecutive \emph{non-catalyzed} reactions, as in Figure 1 of R\"ossler's paper:
\begin{eqnarray}
    &\ch{A + B -> C + D} \label{eq:A}\\
    &\ch{D + E -> F + G} \label{eq:D}\\
    &\ch{G + H -> I + A} \label{eq:G}
\end{eqnarray}
R\"ossler calls this phenomenon \enquote{generalized catalysis} (today we would say \enquote{network catalysis}) and refers to the traditional version as \enquote{true} catalysis. 

In generalized catalysis, the catalyst is not a single molecule, but a set of molecules $\{$\ch{A,D,G}$\}$ that are cyclically transformed into each other while converting input substances $\{$\ch{B,E,H}$\}$ into output substances $\{$\ch{C,F,I}$\}$. At a fine-enough grain of temporal resolution a cyclical scheme must also hold for the process of \enquote{true} catalysis by a single molecule or that molecule would not be restored. Yet, time scales matter: If intermediates are long-lived rather than short-lived, i.e., if they are stable molecules, they can participate in a network of additional reactions. This becomes especially consequential if one of the outputs of the cycle leads through a distinct sequence of reactions back to another member of the cycle (for example, if reaction \eqref{eq:G} directly produced \ch{D} instead of \ch{I} or if \ch{I} reacted further to eventually produce \ch{D}). In that case, all intermediates can be amplified exponentially until input substances are exhausted. R\"ossler encourages chemists to think like engineers who construct complex machinery by combining simple building blocks. For example, adding \ch{E} and \ch{H} as starting materials to the reaction mixture \eqref{eq:A} unlocks \eqref{eq:D} and \eqref{eq:G} and, hence, network catalytic or autocatalytic behavior.

Eigen's 1971 paper deals mainly with autocatalysis as direct self-replication, like in a reaction of the form \ch{A + X -> 2 X + B}, which in R\"ossler's terminology is \enquote{true} autocatalysis. Eigen's model emphasizes macromolecules, such as nucleic acid sequences, that contain \emph{information} for their own replication. The decoding of that information by actual means of chemical production is abstracted away, yielding an appearance of \enquote{true} autocatalysis with associated phenomenological equations. Some of the popularity of Eigen's approach might be due to being firmly situated in a familiar Darwinian framework: Sequences are individualized entities and copying implements imperfect inheritance by descent. In contrast, a network (auto)catalyst is not an individualized entity---we cannot count discrete instances of a chemical cycle, only instances of its components. Network autocatalysis is an emergent organizational pattern, which is why R\"ossler refers to it as \emph{ordinary autonomous growth}. 

Eigen briefly considers network autocatalysis, but in the specific context of polypeptides that catalyze the joining of other polypeptides, which, given reasonable assumptions, can yield cyclical reaction patterns of the catalytic and autocatalytic kind. This scenario, however, is dismissed, because the mechanism is too unspecific to be capable of inheritance by descent and hence of capturing selective advantages that might arise from variation. Kauffman entertains the same idea as Eigen---networks of polypeptides mutually catalyzing their ligation---but approaches it through the lens of a random Boolean network, in which a reaction is a Boolean function (go/no-go of the ligation) controlled by a Boolean input (presence/absence of a suitable polypeptide catalyst). Like Eigen, Kauffman concludes that autocatalytic closure has a meaningful probability of occurrence, but unlike Eigen, Kauffman has no urge to quickly reach the Darwinian shoreline. He seeks to explain patterns of functional order as originating through a process of self-organization distinct from the process through which selection acts upon these patterns to shape surviving forms.

Like for Kauffman, R\"ossler's objective is an appreciation of what can be reasonably expected to occur without, and hence prior to, Darwinian selection. But distinct from Eigen and Kauffman, R\"ossler's contribution is an analysis of what the other two take for granted---catalysis---which leads him to \emph{emergent catalysis} as a cyclical closure of non-catalyzed reactions invoking generic chemistry without appeal to macromolecules. Note that the \enquote{autocatalytic set} of Kauffman requires more than cyclical closure; it requires that the catalysts of each reaction be themselves products of the reaction network. By taking as basic components non-catalyzed reactions, R\"ossler's networks are more primitive and might play a more plausible role in the early stages of chemical evolution on a planet. 

The concept of a catalytic cycle has long been known through the discoveries of the Krebs cycle or the Calvin cycle in the 1930s and 1950s. Although the reactions of these evolved biological cycles are themselves catalyzed, the notion that the cyclical arrangement is also catalytic on its own was well appreciated. At a more basic level, R\"ossler points out that it was the Soviet chemist Nikolai Semenov who discovered in the 1930s \enquote{\emph{isothermal} chemical explosions} \cite{Semenov1959} that keep going by virtue of supply feedback from chemical reactions rather than heat, which is the usual autocatalyst in many \emph{physical} explosions.

R\"ossler considers a chemistry that is abstract, i.e., whose molecules are represented by placeholder variables (such as \ch{A}, \ch{B}, etc) devoid of chemical structure. This is in stark contrast to approaches based on specific flavors of chemistry, such as polypeptides, polynucleotides, iron-sulfur chemistry, lipid or sugar chemistry \cite{mor16a}. While the lack of chemical detail requires making assumptions that can be debated, it permits the formulation of a broad \emph{principle} of chemical evolution.

Central to his framework is a purely conceptual \emph{network of spontaneously possible reactions}, which we abbreviate as NSPR. It is of unfathomable extension, as it consists of all molecules and their reactions provided the latter are both thermodynamically possible (i.e., associated with a decrease in free energy) and kinetically feasible (i.e., having a free energy barrier that can be crossed with possibly low but still meaningful probability). R\"ossler argues that the NSPR contains a very large number of catalytic and autocatalytic cycles.

\begin{figure}
    \centering
    \includegraphics[width=0.6\textwidth]{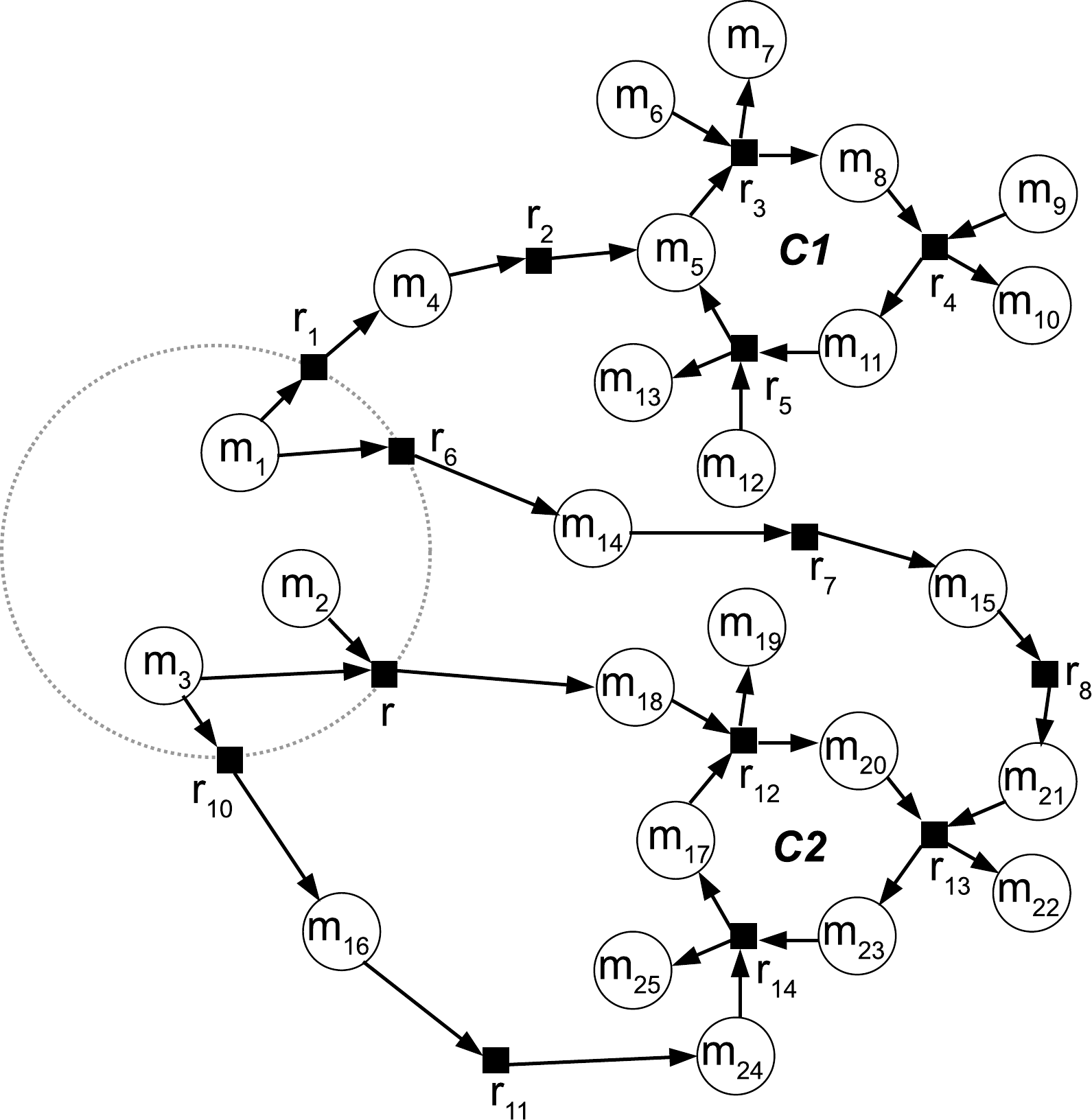}
    \caption{The schematic represents a reaction network in which black squares are reaction nodes and open circles are molecular species. The network contains two catalytic cycles C1 and C2 coupled to the starting set (molecules m$_1$, m$_2$, and m$_3$). In C1, only a member of the catalytic cycle can be produced from the starting set. If all substrates were available, this would suffice to start the cycle, but they are not, because they are not reachable from the starting set. In C2, only the substrate molecules of the cycle can be produced from the starting set. For a catalytic cycle to operate efficiently, both of these conditions must be met. For autocatalytic cycles, the second condition is the most important; it also is the harder one to achieve.}
    \label{fig:paths}
\end{figure}

Assume some tiny set of spontaneously reacting molecules that are initially available in large supply; call them the \enquote{starting set}. The starting set spontaneously initiates an exploration of chemical space. The majority of reactions among the molecules in the starting set are likely to be very slow, causing the exploration to advance initially by single-molecule amounts. This is a regime that mandates a stochastic description, in which a rate constant is the parameter of a probability distribution of time delays to the next reaction event. High free energy barriers mean low rate constants and hence long average time delays. Molecular species that are produced by a sequence of reactions from the starting set become available for further reactions, including loss. After a long time lag, this exploratory process may (or may not) discover one of the great many autocatalytic cycles embedded in the NSPR. R\"ossler emphasizes that the challenge is not the discovery of one member of a cycle by a sequence of reactions from the starting set, but rather the discovery of a cycle for which \emph{all} substrates can be provided in sufficient concentrations. This point is illustrated schematically in Figure \ref{fig:paths}. 

Suppose that, after a possibly long delay, an autocatalytic cycle is actually discovered. This cycle will concentrate mass into its members and its immediate products, which become available in abundances comparable to those of the original starting set. This effectively enlarges the starting set, if it is continuously supplied. The number of possible spontaneous reactions within reach of the exploratory process increases nonlinearly with the \emph{sustained} diversity of molecular species (a network effect) and so does the probability of discovering a further cycle. But any further cycle contributes to sustainably expanding that diversity. Hence there must be a threshold number of discovered cycles beyond which the probability of discovering another cycle approaches 1. From then on, the process of discovering autocatalytic cycles has itself become autocatalytic. R\"ossler calls this behavior \emph{second-order autonomous growth}.

This, then, is the thread through R\"ossler's paper: (i) Catalysis without catalysis: emergent catalysis by cyclic arrangement of non-catalyzed reactions. (ii) First-order autonomous growth: generalized autocatalysis through positive chemical feedback when a cycle directly or indirectly generates one of its members with stoichiometry greater than 1. (iii) Turning the wheels: the discovery of an autocatalytic cycle enables the further exploration of chemical space by effectively extending the reservoir of starting substances. (iv) Second-order autonomous growth: positive feedback on the likelihood of discovering a further cycle with each discovered cycle.

Whether chemistry is a R\"osslerian system for discovering systems, depends on many assumptions. Assessing these assumptions using abstract models informed by empirical data whips up many open questions. What is required for an abstract NSPR to be \enquote{chemical}? Must every reaction be individually \enquote{thermodynamically possible} or should the NSPR include energy transduction? What can we say about the density of cycles in a chemical NSPR? Is the cycle density uniform or does it depend dramatically on specific flavors of chemistry? What are the kinetic requirements that support a given extent of chemical space exploration? In particular, what is required for an autocatalytic cycle to persist in a scenario of open chemical space exploration? A constructively cautionary note by Leslie Orgel is also worth heeding \cite{Orgel2008}.

Chemistry is a bridgehead to life. R\"ossler suggests that to jump start a biology, his process of chemical exploration must discover an autocatalytic system from which further discovery proceeds in less blind a fashion. G\'anti's chemoton comes to mind as a possible specification for when R\"ossler's process morphs into a Darwinian one and life becomes a \emph{controlled} isothermal chemical explosion.

\vspace*{0.2cm}
\subsection*{Acknowledgments}
\vspace*{0.4cm}

We thank James Griesemer for helpful comments. P.~H.~acknowledges support from the Austrian Science Fund (FWF), project J-4537.

\phantomsection
\addcontentsline{toc}{section}{References}
\putbib[refs1]
\end{bibunit}

\begin{thebibliography}{10}
\bibitem{ref_2}
For reversible reaction steps more than one molecule may be necessary. {I}n
  practice, the presence of trace impurities able to react with cycle species
  necessitates using enough starting concentrations or -pressures to saturate
  these impurites.
\bibitem{ref_7}
The number of possible derivates is smaller than the number of macromolecular
  combinations producible by permutation of the starting materials. {H}owever,
  some larger units are almost completely freely permutable. {T}his means that
  the set of all possible products can be compared to a ''small universal
  library". {F}ollowing {K}urd {L}asswitz, an universal library is the set of
  all books producible by permutation of a given alphabet with a given (in this
  case small) length.
\bibitem{ref_9}
When creating a series of several reactions of this type, the time delay does
  not increase in an additive order, but in a multiplicative one.
\bibitem{ref_10a}
In the purely well-mixed case, the final state of the full system could be
  characterized by oscillations of the {V}olterra-{K}erner type \cite{ref_10}.
\end{thebibliography}
\end{document}